# Anomalous roughening of forced radial imbibition in a porous medium

Yong-Jun Chen [1, 2, *], Shun Watanabe [2], Kenichi Yoshikawa [2, *]

[1] Department of Physics, Shaoxing University, Shaoxing, Zhejiang Province 312000, China

[2] Faculty of Life and Medical Sciences, Doshisha University, Kyotanabe, Kyoto 610-394, Japan

∗Corresponding authors

Email address: chenyongjun@usx.edu.cn and keyoshik@mail.doshisha.ac.jp

**Abstract**

We report forced radial imbibition of water in a porous medium in a Hele-Shaw cell. Washburn's law is confirmed in our experiment. Radial imbibition follows scaling dynamics and shows anomalous roughening dynamics when the front invades the porous medium. The roughening dynamics depend on the flow rate of the injected fluid. The growth exponents increase linearly with an increase in the flow rate while the roughness exponents decrease with an increase in the flow rate. Roughening dynamics of radial imbibition is markedly different from one dimensional imbibition with a planar interface window. Such difference caused by geometric change suggests that "universality class" for the interface growth is not universal.

## 1. Introduction

Interface growth is found throughout nature [1]. Propagation of an interface shows roughening dynamics [1]. Roughening dynamics of the interface are characterized by sets of critical exponents. Different interface growths with the same critical exponents belong to the same universal class of roughening dynamics [2]. The roughening process of a propagating front is often described in terms of the root-mean-square width $W$ of the interface fluctuation. Family and Vicsek (FV) proposed a dynamic scaling hypothesis for a fixed-size interface [1, 3]: $W(L) \sim t^\beta$ for $t < \tau$, $W(L) \sim L^\alpha$ for $t > \tau$ and $\tau \sim L^z$, where $L$ is the lateral size of the interface, $t$ is time, $\tau$ is the saturation time of roughness, $\beta$ is a growth exponent, $\alpha$ is a roughness exponent, and $z = \alpha/\beta$ is a dynamic exponent. The scaling dynamics of the initially planar interface with a fixed-size window have been studied intensively [1]. In addition, the role of geometric properties, for example, the size of the interface window, the shape of the interface, and non-Euclidean geometry, has attracted attention [4-6]. As a matter of fact, many systems do not exhibit a fixed-size window during interface propagation, including both biological and physical interfaces. In these cases, the interface grows in all directions, and forms not only overhangs but also non-Euclidean geometry. Among these, a traditional scaling analysis has been adopted to understand the universality class of tumor growth [7]. While experimental studies have shown that both the radial and linear growth of a cell colony front obey the universality class of the Kardar-Parisi-Zhang (KPZ) model [8, 9], the relation between the scaling behaviors of fixed-size growth and radial growth is still under debate. In this article, we observed radial imbibition in a porous medium confined within a Hele-Shaw cell, as a simple and representative model system for studying various growth phenomena.

Imbibition, or the displacement of one fluid by another in a porous medium, exhibits a roughening transition with scaling dynamics during the motion of the invading front [10]. The kinetic roughening process is governed by non-local dynamics. Due to its importance in industry, such as in oil recovery, printing, food industry, textile construction and medicine transport, and to gain a fundamental understanding of a nonequilibrium process, imbibition has been studied intensively both theoretically and experimentally [11-21]. For a fixed-size interface window, the anomalous roughening of imbibition, where local exponents are substantially different from global exponents, has recently been observed [11-13]. It is believed that roughening dynamics are caused by the disorder in the field of the porous medium, including capillary disorder, permeability disorder, pressure disorder, and so on [10]. The local motion of the interface is related to the curvature of the local interface, and is governed by Darcy's law [22, 23]. Thus, it is reasonable to adopt a curvature-driven model to describe the motion of the imbibition front [24]. However, recent studies have shown that geometric differences of propagation window can cause a striking difference in the scaling dynamics [5, 6]. Here, we observed radial forced fluid imbibition in a porous medium. We found that radial imbibition follows scaling dynamics and the scaling behavior depends on the flow rate of the fluid. Anomalous roughening behavior was also found in radial forced imbibition. In comparison with one-dimensional imbibition with a fixed-size window, radial imbibition in two dimensions demonstrates new roughening dynamics.

## 2. Experiment

The experiment was performed using a Hele-Shaw cell constructed from two glass plates (30cm×30cm) aligned parallel to each other with a gap size of 0.5mm. The Hele-Shaw cell was filled with approximately two layers of glass beads with a diameter of 0.20±0.05mm. The glass beads were packed tightly by manual vibration during filling. Three of the four boundaries of the Hele-Shaw cell were sealed while one was left open so that air could come out. Pure water was injected from the center of the cell at constant flow rates between 0.060ml/min and 2.0ml/min, using an auto syringe pump. The invasion under a fixed injection flow rate was observed repeatedly and its statistical reproducibility was confirmed. The invasion of the imbibition front was monitored using a digital CCD camera. Lateral optical illumination yielded a sharp contrast at the invading front. The experimental data were analyzed using image-analysis software. All experiments were performed under ambient conditions (20℃).

## 3. Results and discussion

Figure 1(a) shows a typical radially propagating front of an air-water interface. The front roughens during invasion into the air in the porous medium as shown in Fig. 1(b). The average position $\bar{r}$ of the front, i.e., the radius to the center of mass of the front, evolves according to Washburn's law, $\bar{r} = At^{1/2}$, where $t$ is time and $A$ is a constant, as shown in Fig. 1(c). Conservation of the liquid volume leads to slowing of the average position of front. The flow in the porous medium can be described using Darcy's law, $d\bar{r}/dt \sim \nabla p \sim f/\bar{r}$, where $p$ and $f$ are the pressure and the injection flow rate. Consistent with Washburn's law, $A^2$ increases linearly with an increase in the flow rate, $A^2 \sim f$, as shown in the inset of Fig. 1(c). For a large flow rate, as $f = 2000 \mu l/\min$, the increase in $A^2$ decreases relative to the linearity (inset in Fig. 1(c)). To our knowledge, this is the first report to confirm Washburn's law in radial imbibition.

In our experiment, the front was digitized and analyzed numerically. Before calculating the roughness of the interface, we redistributed the original digitized points to uniform distribution on the front interface. The overhangs were included in our calculation. The roughness of the front is the root-mean-square fluctuation

$$W = \langle r^2 - \bar{r}^2 \rangle^{1/2},$$

where $r$ and $\bar{r}$ are the radius to the center of mass (CM) and its mean value, and $\langle \rangle$ represents the average over the space. Figure 2 is the log-log plot of roughness against time. The evolution of roughness exhibits scaling dynamics of interface propagation. The fluctuation in the log-log plot of the scaling law is attributable to the overhang and coalescence of the front. We obtained growth exponents of 0.26, 0.28, 0.38, 0.56, and 0.71 for flow rates of 60, 200, 400, 1000, and 2000 μl/min, respectively. The growth exponent increases linearly with the flow rate, as shown in the inset in Fig. 2. In a similar study, the dependence of the growth exponent on the flow rate has been reported in the spontaneous imbibition of water into a controlled gap disorder in a Hele-Shaw cell with a fixed-size window [13], which shows that the invasion front exhibits different roughening exponents under different pressures. Those authors reported that the growth exponent decreases when the pressure increases under a larger flow rate, which is the inverse of our result. Roughness was not saturated in our experiments. For circular propagation, roughness should usually show saturation because the curvature decreases with an increasing in the mean radius of the front [6]. This suggests that the correlation length of the interface grows slowly and is always smaller than the size of the invasion front. This is markedly different from the scaling behavior in the case of a fixed-size interface planar window [1, 25].

To obtain the roughness exponent, we calculated the structure function of the invading front, $S(k,t) = |r(k,t)|^2$, where $r(k,t)$ is a Fourier transformation of the radius $r$ and $k$ is the wave number. For a self-affine interface, the structure function has the scaling form [2]

$$S(k,t) = k^{-(2\alpha+1)} f_s(kt^{1/z}) \text{ with}$$

$$f_s(u) \sim \begin{cases} u^{2(\alpha-\alpha_s)} & \text{if } u \gg 1 \\ u^{2\alpha+1} & \text{if } u \ll 1 \end{cases}$$

where $f_s(u)$ is a spectral scaling function, $\alpha$ is a roughness exponent, and $\alpha_s$ is a spectral roughness exponent. Figure 3 shows the log-log plot of the structure function against wavenumber. Figure 3a is the structure function at different time points with a flow rate of 200μl/min. There are two regimes with different scaling behaviors, i.e., a small value of $k$ (corresponding to a large size in real space) and a relatively large value of $k$ (corresponding to a small size in real space). The roughness exponent does not change with time, as shown in Fig. 3a. The inset in Fig. 3a shows a log-log plot of $S(k,t)k^{(2\alpha+1)}$ versus $kt^{1/z}$ with roughness exponent $\alpha = 1.2$ when the flow rate is 200 μl/min. The spectral scaling function $f_s(kt^{1/z})$ collapses to the same scaling at a different time. From the collapse of the spectral scaling function, we found that $\alpha = \alpha_s$. As shown in Fig. 3(b), the roughness exponent is 1.30±0.1, 1.20±0.1, 1.10±0.1, 0.95±0.1 and 0.90±0.1 for flow rates of 60 μl/min, 200 μl/min, 400 μl/min, 1000 μl/min, and 2000 μl/min, respectively. From the surface geometry, it is easy to see the general trend that the front is rougher and thus has a greater roughness exponent when the flow rate is smaller.

The local roughening dynamics of the invasion front were calculated. We divided the closed front into 12 equal parts using the center angle from the mass center of the front, as shown in Fig.

4. The local roughness of the front and the local structure function are $W_{loc}(\theta,t) = \langle r^2 - \bar{r}^2 \rangle_\theta^{1/2}$ and $S_{loc}(k,t) = |r(k,t)|_\theta^2$, respectively. Figure 4 shows the typical log-log plots of the local roughness against time (Fig. 4(a)) and the local structure function against wave number (Fig. 4(b)) for invasion fronts of different sizes (center angle from the mass center) when the flow rate is 200μl/min. The roughness evolution and structure function depend on the center angle of the interface. According to our calculation, the local growth exponents $\beta_{loc}$ are 0.16±0.05, 0.16±0.05, 0.24±0.05, 0.40±0.05, and 0.50±0.05 for the flow rates of 60 μl/min, 200 μl/min, 400 μl/min, 1000 μl/min, and 2000 μl/min, respectively. And the local roughness exponents $\alpha_{loc}$ are 1.0, 1.0, 1.0, 0.7 and 0.5 for flow rates of 60 μl/min, 200 μl/min, 400 μl/min, 1000 μl/min, and 2000 μl/min, respectively. If we collapse the data in Fig. 3(a) (the inset of Fig. 3(a)), we have $\alpha_s = \alpha$. When $\alpha > 1.0$ and $\alpha_{loc} = 1.0$, the interface is super-rough with flow rates of 60 μl/min, 200 μl/min, and 400 μl/min. Such scaling behavior is similar to the observation with a fixed-size window when the pressure is negative [12, 13].

Let us compare the above results with those in forced flow imbibition (FFI) with a fixed-size window in quenched disorder. As reported by Soriano *et al.* [25], the growth exponent of FFI with a fixed-size window is 0.5, which is independent of the experimental parameters, and the roughness exponent on a short length scale decreases with a decrease in the flow rate. For weak quenched disorder, the scaling behavior does not depend on the disorder configuration [25]. However, for strong quenched disorder, FFI demonstrates anomalous roughening behavior with $\beta = 0.50$, $\beta_{loc} = 0.25$, $\alpha = 1.0$, and $\alpha_{loc} = 0.5$ when the flow rate is smaller than a threshold flow rate [26]. There is a clear difference between fixed-sized FFI and our radial FFI, as shown above. In our radial FFI, due to the enlargement of the fluid front and the conservation of liquid, the local flow rate decreases in the radial direction. Such different scaling dynamics is related not only to the different disorder but also to the different geometry of the interface.

For imbibition with a fixed-size planar window, the finite size of the interface window and the boundary effect should affect the observed scaling behavior. As known in anomalous roughening, a different size for the interface should lead to a difference in the local roughness evolution and structure function [1]. Especially, the boundary effects on local roughening dynamics are critical, and yet still not clear [10]. As shown in an experiment [10], the boundary will tilt the invading front, and we have to subtract such tilting caused by the boundary effect. However, this subtraction of the interface morphology must be performed very carefully according to the insight obtained in our present study. If we include boundary effects, local disorder and so on, we obtain exponents with large variation [10]. For radial imbibition, while there is no such disadvantage we do observe enlargement of the interface. In addition, a previous consideration of the interface roughening mainly focused on one-dimensional propagation of the interface, while lateral "soliton-like" propagation was neglected [27]. In our radial imbibition, lateral propagation is nontrivial. For local roughening dynamics, interface growth should follow the same physical process for both fixed-size and radial imbibition. As shown by the simulation based on curvature-driven growth of interface [6], a variable interface window leads to striking difference in the roughening dynamics of the interface propagation.

## 4. Conclusion

In summary, we observed the scaling dynamics of radial imbibition in a porous medium made from close-packed glass beads in a Hele-Shaw cell. We demonstrate the first confirmation of Washburn's law in radial imbibition. Radial imbibition exhibits anomalous roughening dynamics. The growth and roughness exponents depend on the flow rate of the injected fluid. The local exponents are substantially different from the global exponents. The imbibition front is super-rough when the flow rate is small because the pinning effect dominates the roughening of the invading front. The radial imbibition in our porous medium is strikingly different from that with a fixed-size interface window. These findings provide the new horizon to understand the essential difference between the scaling behaviors of fluctuation in interface propagation with a fixed-size window and a variable window. As a final remark, we would like to point out that we cannot assume in advance that the geometric shape of interface will not affect the roughening dynamics of the propagating interface. A naive imagination on the roughening dynamics of circular KPZ growth will shows that the values of growth exponent and roughening exponent do not depend on the shape of interface and belong to the same KPZ universality class of one-dimensional planar growth. The interface growing which belongs to the same universality class will have same mechanism of interface growth [1, 2]. As made clear in the present article, roughening dynamics of the radial propagation of interface is different from one-dimensional roughening process and has different exponents. Thus, we should think more deeply whether the same class of interface growth with different interface shapes belongs to the same class of roughening dynamics, and, more specifically, whether circular "KPZ growth" and one-dimensional "KPZ growth" belong to the same KPZ universality class (with different exponents). If they do not belong to the same class because of different exponents, what does the term "universality class" mean or does "universality class" really exist? Physically, change from one-dimensional growth to circular growth should NOT lead to change of the general mechanism of the interface growing, which determines the exponents of interface roughening. Maybe we could conclude that the "universality class" for the interface growth does not exist. We have immersed ourselves in studies on the interface growth and the celebrated KPZ universality class for over two decades. Shall we continue to find universality class of the interface growth in nature? We hope that the above thought will stimulate new direction of studies on the interface growth.

**Acknowledgements**

This work was supported by the National Natural Science Foundation of China (No. 11204181), SRF for ROCS, SEM and Kakenhi (Nos. 23240044, 25103012).

**Figure captions**

Figure 1 Imbibition in close-packed glass beads. (a) A typical morphology of imbibition front. Scale bar: 50mm. (b) Spatial-temporal evolution of the invasion front. The time interval between neighboring fronts is 16.67s. The flow rate is 200μl/min. (b) Plot of the squared mean radius $\bar{r}^2$ against time. The linear fit demonstrates Washburn's law for the imbibition of fluid in a porous medium. The inset is a plot of the squared prefactor $A^2$ in Washburn's law against the flow rate.

Figure 2 Log-log plot of roughness against time at different flow rates. The guidelines of linear fit from up to down have slopes of 0.26, 0.28, 0.38, 0.56, and 0.71, respectively. To avoid crowding of the data, -1, -0.5, and -0.25 have been added to the logarithm of roughness for flow rates of 2000μl/min, 1000μl/min and 400μl/min, respectively. The right inset is a plot of the growth exponent β against the flow rate.

Figure 3 Log-log plot of the structure function $S(k,t)$ against $k$. (a) Structure function against $k$ for a flow rate of 200 μl/min at different time points. Inset: scaling function $f(kt^{1/z}) = S(k,t)k^{2\alpha+1}$ vs $kt^{1/z}$ ($\alpha = 1.2$, $z = \beta/\alpha = 0.23$). (b) Structure function against $k$ for different flow rates (unit: μl/min). The data were artificially shifted vertically. The dashed lines are guidelines for the data and the slope ($-(2\alpha+1)$) is indicated in the figure.

Figure 4 Log-log plot of interface roughness and structure function for different window sizes. (a) Local roughness of the invasion front as a function of time. The slopes of dashed guidelines from up to down (center angle from $2\pi$ to $\pi/6$) are 0.28, 0.27 0.25, 0.24, 0.24, 0.23, 0.21, 0.16, 0.16, 0.16, 0.16, and 0.16, respectively. We have $\beta_{loc} = 0.16$. (b) Local structure function. The time is 86.67s. The slopes of the dashed guidelines are -3.4, -3.3,-3.2,-3.1,-3.0, -3.0 and -3.0, respectively. We have $\alpha_{loc} = 1.0$. The flow rate is 200 μl/min. The data was artificially shifted vertically.

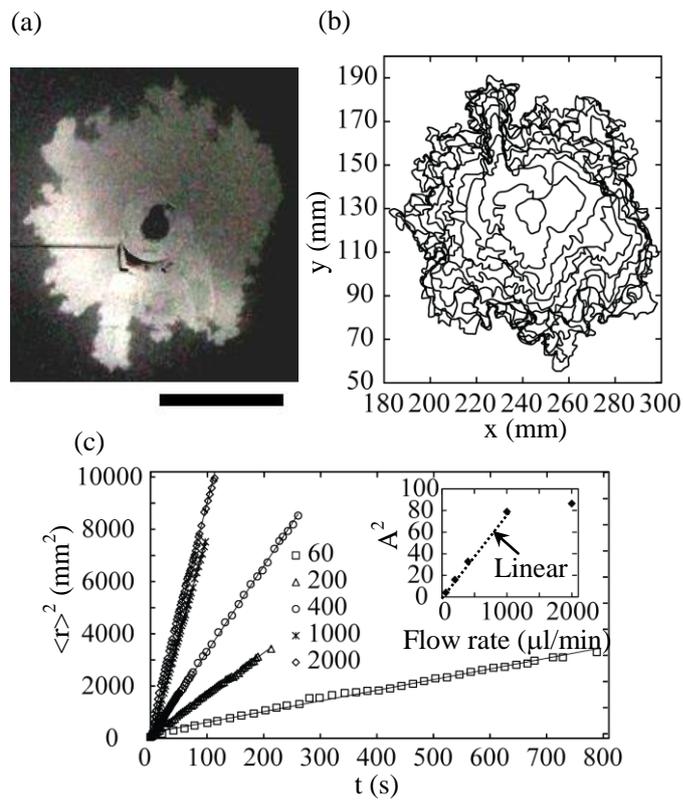

Figure 1 Yong-Jun Chen, Shun Watanabe, Kenichi Yoshikawa

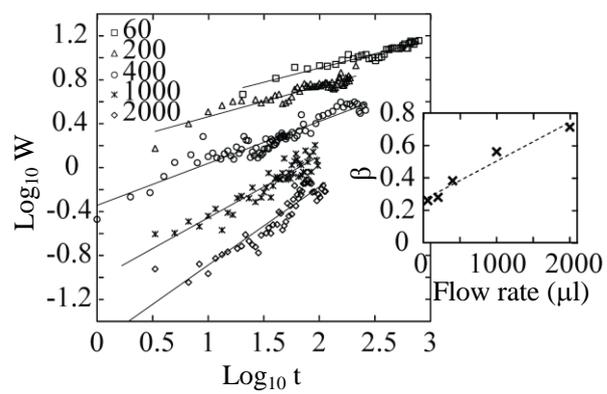

Figure 2 Yong-Jun Chen, Shun Watanabe, Kenichi Yoshikawa

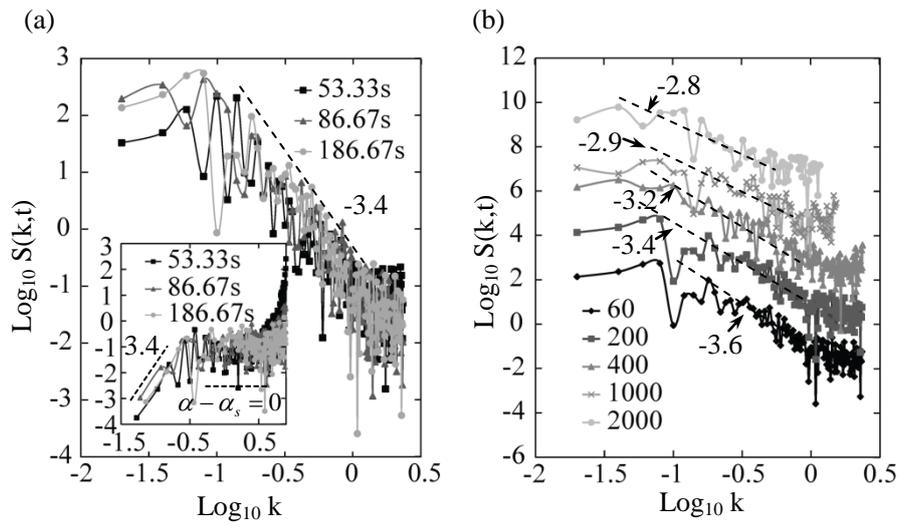

Figure 3 Yong-Jun Chen, Shun Watanabe, Kenichi Yoshikawa

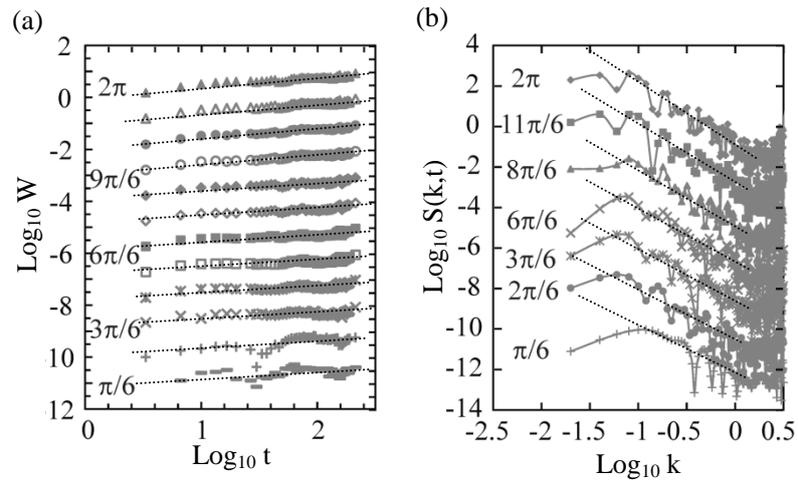

Figure 4 Yong-Jun Chen, Shun Watanabe, Kenichi Yoshikawa